\documentclass[aps,twocolumn,showpacs,floats,nofootinbib,preprintnumbers]{revtex4}
\usepackage{graphicx,color,natbib}
\usepackage{amsmath,amssymb,amsfonts}
\newcommand{\bse}{\begin{subequations}}
\newcommand{\ese}{\end{subequations}}
\newcommand{\be}{\begin{equation}}
\newcommand{\ee}{\end{equation}}
\newcommand{\bea}{\begin{eqnarray}}
\newcommand{\eea}{\end{eqnarray}}
\newcommand{\ba}{\begin{array}}
\newcommand{\ea}{\end{array}}

\input amssym.def
\input amssym.tex

\usepackage[colorlinks=true, linkcolor=blue, bookmarks=true]{hyperref}

\begin{document}
\title{ Holographic Mutual and Tripartite  Information\\
   in a Non-Conformal Background }
\vspace{1cm}
\author{M. Ali-Akbari\footnote{$\rm{m}_{-}$aliakbari@sbu.ac.ir}}
\affiliation{Department of Physics, Shahid Beheshti University G.C., Evin, Tehran 19839, Iran}
\author{M. Asadi\footnote{$\rm{m}_{-}$asadi@ipm.ir}}
\affiliation{School of Physics, Institute for Research in Fundamental Sciences (IPM),
P.O.Box 19395-5531, Tehran, Iran}
\author{M. Rahimi\footnote{$\rm{me}_{-}$rahimi@sbu.ac.ir}}
\affiliation{Department of Physics, Shahid Beheshti University G.C., Evin, Tehran 19839, Iran}

\begin{abstract}
Holographic mutual and tripartite information have been studied in a non-conformal background. We have investigated how these observables behave as the energy scale and number of degrees of freedom vary. We have found out that the effect of degrees of freedom and energy scale is opposite. Moreover, it has been observed that the disentangling transition occurs at large distance between subregions in non-conformal field theory independent of $l$. The mutual information in a non-conformal background remains also monogamous.
\end{abstract}
\maketitle
\tableofcontents
%
\section{Introduction}
The gauge/gravity duality provides a significant framework to study key properties of the boundary field theory dual to some gravitational theory on the bulk side \cite{Maldacena}. The most concrete example of gauge/gravity duality is the $Anti-desitter/Conformal$  correspondence which proposes a duality between asymptotically $AdS$ spacetimes in $d+1$ dimensions and $d-$dimensional conformal field theories. This outstanding correspondance is indeed a strong-weak duality which makes it possible to study various aspects of the strongly coupled systems such as quantum chromodynamic, quark-qluon plasma and condenced matter \cite{CasalderreySolana:2011us,Camilo:2016kxq,Amiri-Sharifi:2016uso}. Although, study of different observables may not seem simple in the feild theory side the duality indeed proposes applicable prescription in the gravity side which makes the calculations as much as simple.

The applicability of the gauge/gravity duality is not restricted to $CFT$'s. It is then important to develop our understanding of this duality for more general cases. There are many different families of non-conformal theories which one can study the effect of the non-coformality on their physical quantities \cite{Attems:2016ugt,Pang:2015lka}.

When one studies properties of a given quantum field theory, it is common to  investigate behaviors of correlation functions of local operators in the theory. However, properties of non-local quantities are equally important. One example of important non-local physical quantities in field theory with a well known dual gravity description is the entanglement entropy which nicely characterizes quantum entanglement between two subsystems $A$  and  its complement $\bar{A}$ for a given pure state. Since the quantum field theories have infinitely degrees of freedom, the entanglement entropy is divergent. In \cite{Srednicki:1993im}, it has been shown that the leading divergence term is proportional to the area of the entangling surface. In the language of $AdS/CFT$, the entanglement entropy has a  holographic dual given by the area of minimal surface extended in the bulk whose boundary coincides with the boundary of the subregion \cite{Ryu:2006bv,Ryu:2006ef}. The study of entanglement entropy has not also restricted to the $CFT$'s. For instance in \cite{Rahimi:2016bbv,Rahimi:2018ica} the authors  have been used this quantity to probe non-conformal theories nicely.

Due to the $UV$ divergence structure of entanglement entropy, it is better to introduce an appropriate quantity, which is just a linear combination of entanglement entropy and then remain finite, called mutual information. It is an important concept in information theory. For two subregions $A$ and $B$, it is more natural to compute the amount of correlations (both classical and quantum) between these two subregions which is given by the mutual information \cite{Fischler:2012uv,Allais:2011ys}. Note that the subadditivity property of the entanglement entropy gurantees that mutual information is always non-negative. The tripartite information is another important quantity which is defined for three subregions and measures correlation between them. In fact, this quantity can be used to  measures the extensivity of the mutual information. It is also free of divergence and can be positive, negative or zero \cite{Hayden:2011ag,Asadi:2018ijf,Asadi:2018lzr}. 

The background we have considered in this paper is a holographic $5$-dimensional model consisting of Einstein gravity coupled to a scalar field with a non-trivial potential, which is negative and has a minimum and a maximum for finite values of scalar field. Each of these extrema corresponds to an $AdS_5$ solution with different radii\cite{Attems:2016ugt}. In the gauge theory the $4$-dimensional  boundary  is not conformal and, at zero temperature,  flows from an $UV$ fixed point to an $IR$ fixed point. This renormalisation group is dual on the gravity side to a geometry that interpolates between two $AdS$ spaces. We are now interested in study holographic mutual and tripartite information in a non-conformal background and study the effect of field theory parameters such as energy scale and number of degrees of freedom on these quantum information quantities.
 \section{Review}
The holographic model  we study here  is a five-dimensional Einstein gravity coupled to a scalar field with a non-trivial potential whose action is given by
\begin{eqnarray}\label{action}
S=\frac{2}{(G_N^{5})^2}\int d^5x\:\sqrt{-g}\:\big[\frac{1}{4}\mathcal{R}\:-\:\frac{1}{2}(\bigtriangledown\phi)^2\:-\:V(\phi)\big],
\end{eqnarray}
where $G_N^{5}$ is the five-dimensional Newton constant and $\mathcal{R}$ is the Ricci scalar of curvature corresponding to the metric $g$. Scalar field and its potential are also denoted by $\phi$ and $V(\phi)$, respectively.

 In order to have a bottom-up model  the following  potential has been choosen \cite{Attems:2016ugt}
\begin{align}
\begin{split}
L^2V(\phi)&=-3-\frac{3}{2}\phi^2 -\frac{1}{3}\phi^4 + \big(\frac{1}{3\phi_M^2 }+\frac{1}{2\phi_M^4}\big)\phi^6\\
& -\frac{1}{12\phi_M^4}\phi^8 .
\end{split}
\end{align}
This potential possess a maximum at $\phi=0$ and a minimum at $\phi=\phi_M>0$, each of them corresponds to an  $AdS_5$ background with different radii. In the language of the gauge theory, each of these solutions is dual to a fixed point of the $RG$  flow from the $UV$ fixed point at $\phi=0$ to the $IR$ fixed point at $\phi=\phi_M>0$.  It is easy to see that the radii of these asymptotically $AdS_5$ take the form \cite{Attems:2016ugt}
\begin{eqnarray}\label{sAB}
{L_{UV(IR)}=\sqrt{\frac{-3}{V(\phi)}}}=
\begin{cases}
L_{UV}=L & \: \phi=0 , \\
L_{IR}=\frac{L}{1+\frac{\phi_M ^2}{6}} & \: \phi=\phi_M . \\
\end{cases}
\end{eqnarray}
Following the fact that $L_{IR}<L_{UV}=L$ and according to gauge-gravity dictionary, indicated that the number of degrees of freedom in the gauge theory is related to the radius of the background,  a smaller number of degrees of freedom lives in the $IR$ limit, $i.e$. $N_{IR}<N_{UV}$. Furtheremore, as one  increases $\phi_M$ the difference in degrees of freedom between the $UV$ and $IR$  fixed points increases.

If one interested in domain-wall solutions which are interpolating between the two underlying $AdS_5$ backgrounds, the vacuum solutions to the Einstein equations can be obtained from the action (\ref{action}) . The parametrized metric for arbitrary $\phi_M$ can be read
 \begin{eqnarray}\label{metric}
ds^2=e^{2A(r)}(-dt^2+dx^2)+dr^2,
\end{eqnarray}
where 
\begin{align}
e^{2A(r)}&=\frac{\Lambda ^2 L^2}{\phi ^2}(1-\frac{\phi ^2}{\phi _M ^2})^{1+\frac{\phi _M^2}{6}}\,\,\,\,\, e^{-\frac{\phi ^2}{6}},\\
\phi (r)&=\frac{\Lambda L e^{-\frac{r}{L}}}{\sqrt{1+\frac{\Lambda ^2L^2}{\phi _M^2}e^{-\frac{2r}{L}}}},
\end{align}
where $\Lambda$  is the energy scale that break the conformal symmetry in the dual gauge theory. It is also related to the asymptotic value of the scalar field, $i.e$. $\phi(r\rightarrow \infty)$.  For more details about the background see \cite{Attems:2016ugt}.

Entanglement entropy is one of the most significant features of quantum physics and plays a significant role in understanding quantum many-body physics, quantum field theory, quantum information and quantum gravity. Consider a constant time slice in a $d-$dimensional quantum field theory and divide it into two spatial regions $A$ and $\bar{A}$ where they are complement to each other. In quantum field theory, the reduced density matrix for region $A$ can be computed by integrating out the degrees of freedom living in $\bar{A}$, $ i.e$. 
$\rho_A=Tr_{\bar{A}}\, \rho$ where $\rho$ is the total density matrix. The entanglement entropy measures the entanglement between an arbitrary subregion $A$  and its complement $\bar{A}$. It is defined as the von Neumann entropy of the reduced density matrix
\begin{eqnarray}
S_{A}=-tr \rho_{A}\log\rho_{A},
\end{eqnarray}
where $\rho_{A}$ is the reduced density matrix of $A$.
In general, it is difficult to compute entanglement entropy directly due to the infinite degrees of freedom in a field theory. Motivated by this and by applying $AdS/CFT$ correspondence, a holographic prescription, known as $\text{Ryu}$ and $\text{Takayanagi}$ ($RT$) formula,  has been proposed to compute entanglement entropy through the following area law relation \cite{Ryu:2006bv,Ryu:2006ef}
\begin{eqnarray}\label{RT}
S_{A}=\frac{Area(\gamma_{A})}{4G_{N}^{d+2}},
\end{eqnarray}
 where $\gamma_{A}$ is the $d-$dimensional extremal surface in $AdS_{d+2}$  whose boundary is given by $\partial{A}$ and $G_{N}^{d+2}$  is the $d+2-$ dimensional Newton constant. Intrestingly, one can extend this formula  to any asymptotically $AdS$ spaces.
 \begin{figure}[ht] 
\begin{center}
\includegraphics[width=70mm]{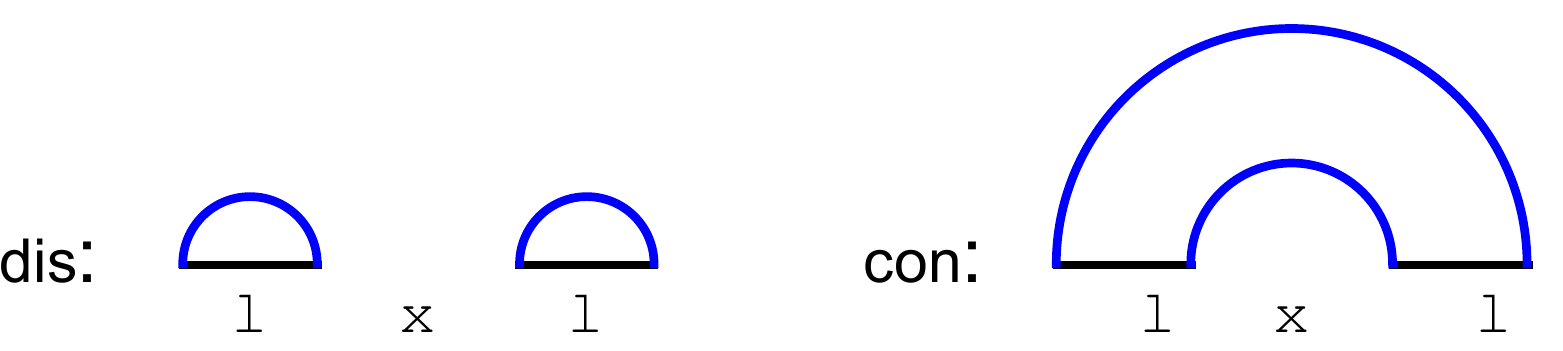} 
\hspace{10mm}
\caption{Two different configurations for computing $S_{A\cup B}$. The time coordinate is suppressed.}
\label{fig:s}
\end{center}
\end{figure}

 When the boundary region we are interested in is made by two disjoint  $A$ and $B$, the most important quantity to consider is the mutual information which measures the amount of correlation between $A$ and $B$ \cite{Fischler:2012uv}
  \begin{eqnarray}
 I(A,B)\equiv S(A)+S(B)-S(A\cup B),
 \end{eqnarray}
    where $S(Y)$ denotes the entanglement entropy of the region $Y$. It is a finite, regulator-independent and non-negative quantity. Note that when two subregions are close enough to each other, there is a finite positive correlation between them. However, when they are very far apart from each other  the mutual information vanishes. The transition of the mutual information from positive values to zero occurs at the distance which we call $x_{DT}$. We consider the holographic mutual information for the metric \eqref{metric}. This quantity depends on many variables and our analysis is mainly numerical. We take equal subregions $A$ and $B$ whose lengths are $l$ and the separation length is $x$.
  
  Given the underlying subregions $A$ and $B$, the entanglement entropy of each subregion can be computed from \eqref{RT}. However, the computation of $S(A\cup B)$ is more interesting. There are two minimal surfaces, which are schematically shown in Fig. \ref{fig:s} , extending in the bulk whose boundaries coincide with$\partial A\cup B$. Thus, $S(A\cup B)$ is  \cite{Asadi:2018ijf}
\begin{eqnarray}\label{sAB}
{S_{A\cup B}}=
\begin{cases}
2 S(l), & x>x_{DT}, \\
S(2l+x) +S(x), & x<x_{DT}, \\
\end{cases}
\end{eqnarray}
where $S(Y)$ denotes the area of the minimal surface whose boundary is coincided with the boundary of the underlying subregion. The holographic mutual information of two disjoint is then given by 
\begin{eqnarray}\label{MI}
{I(A,B)}=
\begin{cases}
0, \,\,\,\,\,\,\,\,\,\,\,\,\,\,\,\,\,\,\,\,\,\,\,\,\,\,\,\,\,\,\,\,\,\,\,\,\,\,\,\,\,\,\,\,\,\,\,\,\,\,\,\,\,\,\,\,\,\,\,\,\,\,\,\,\,\,\,\,\,\,\,\,\,\,x>x_{DT}, \\
2S(l) - S(2l+x) -S(x), \qquad x<x_{DT}. \\
\end{cases}
\end{eqnarray}

\begin{figure}[h]
\begin{center}
{\includegraphics[width=70mm]{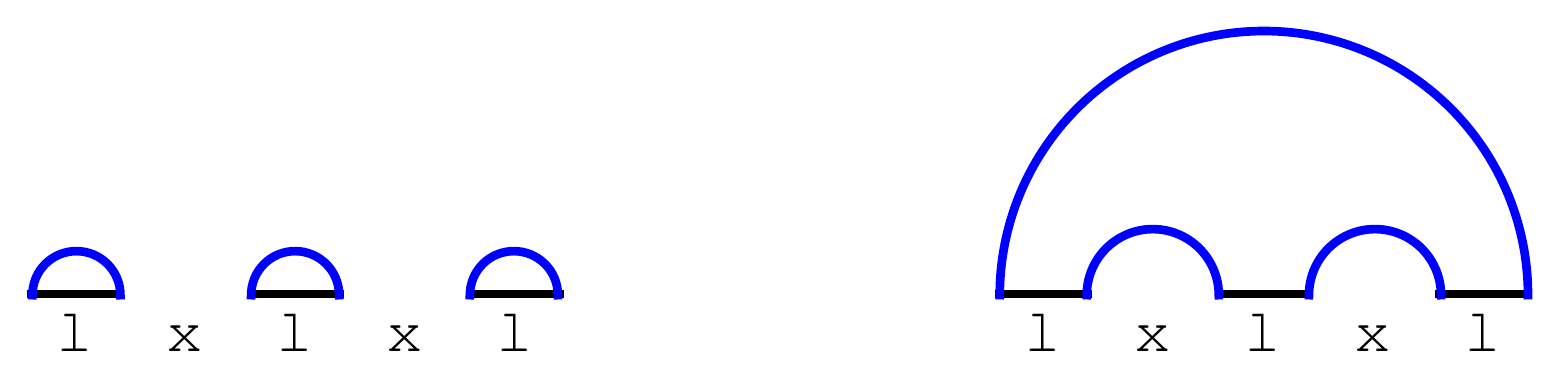}}\\ 
{\includegraphics[width=70mm]{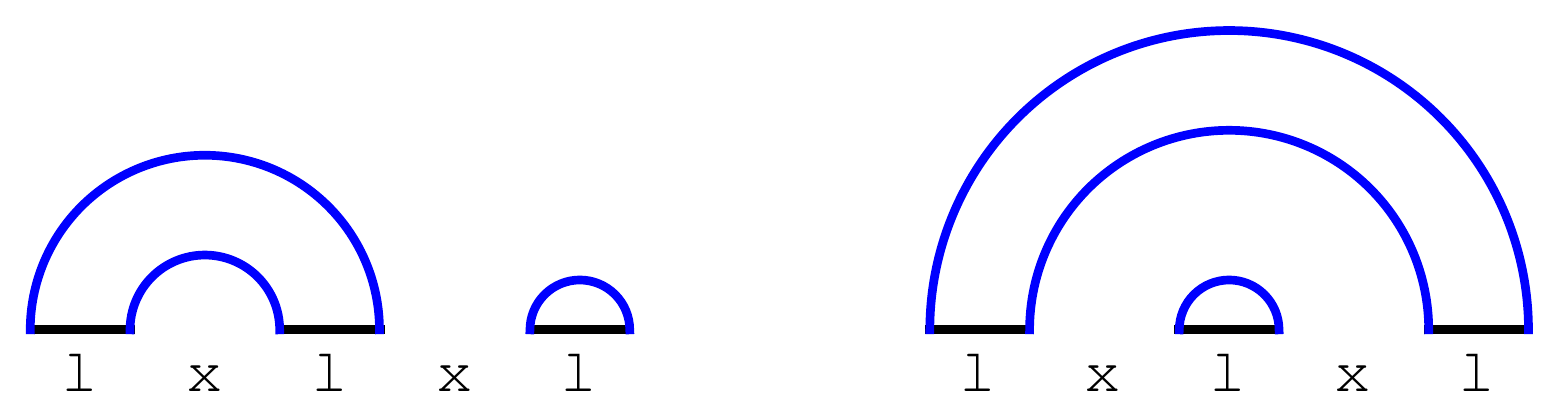} }
\hspace{10mm}
\caption{Four different configurations for computing $S_{A\cup B \cup C}$. The time coordinate is suppressed. }  \label{figI3}
\end{center}
\end{figure}

 Another interesting quantity that can be defined from the entanglement entropy is the tripartite information
\begin{align}\label{I3}
\begin{split}
I^{[3]}(A, B, C)&\equiv S(A)+S(B)+S(C)-S(A\cup B)\\
&-S(A\cup C)-S(B\cup C)+S(A\cup B\cup C)\\
&= I(A,B) + I(A,C)- I(A,B,C)\, ,
\end{split}
\end{align}
where $A$, $B$ and $C$ are three disjoint subsystems. The tripartite information is a measure of the extensivity of the mutual information. According to the $RT$ formula, the mutual information is always extensive or superextensive, $I^{[3]}=0$ and $I^{[3]}<0$ respectively. In either cases the holographic mutual information is called monogamous \cite{Hayden:2011ag}. In addition to the computations that we have already done, we need to compute $S(A\cup B\cup C)$ which is more chalenging. For the union of three subsystems one should consider different configurations for the extremal surfaces Fig. \ref{figI3}.

In \cite{Rahimi:2016bbv} the authors have nicely studied the entanglement entropy between a subregion $A$ and its complement $\bar{A}$, living in the boundary of the metric \eqref{metric}.

In this paper we have considered boundary subregions  described by the background \eqref{metric} and
 the effect of the energy scale $\Lambda$ and the number of degree of freedom on the mutual and tripartite information have been studied.

 \begin{figure}[ht]
\begin{center}
\includegraphics[width=70mm]{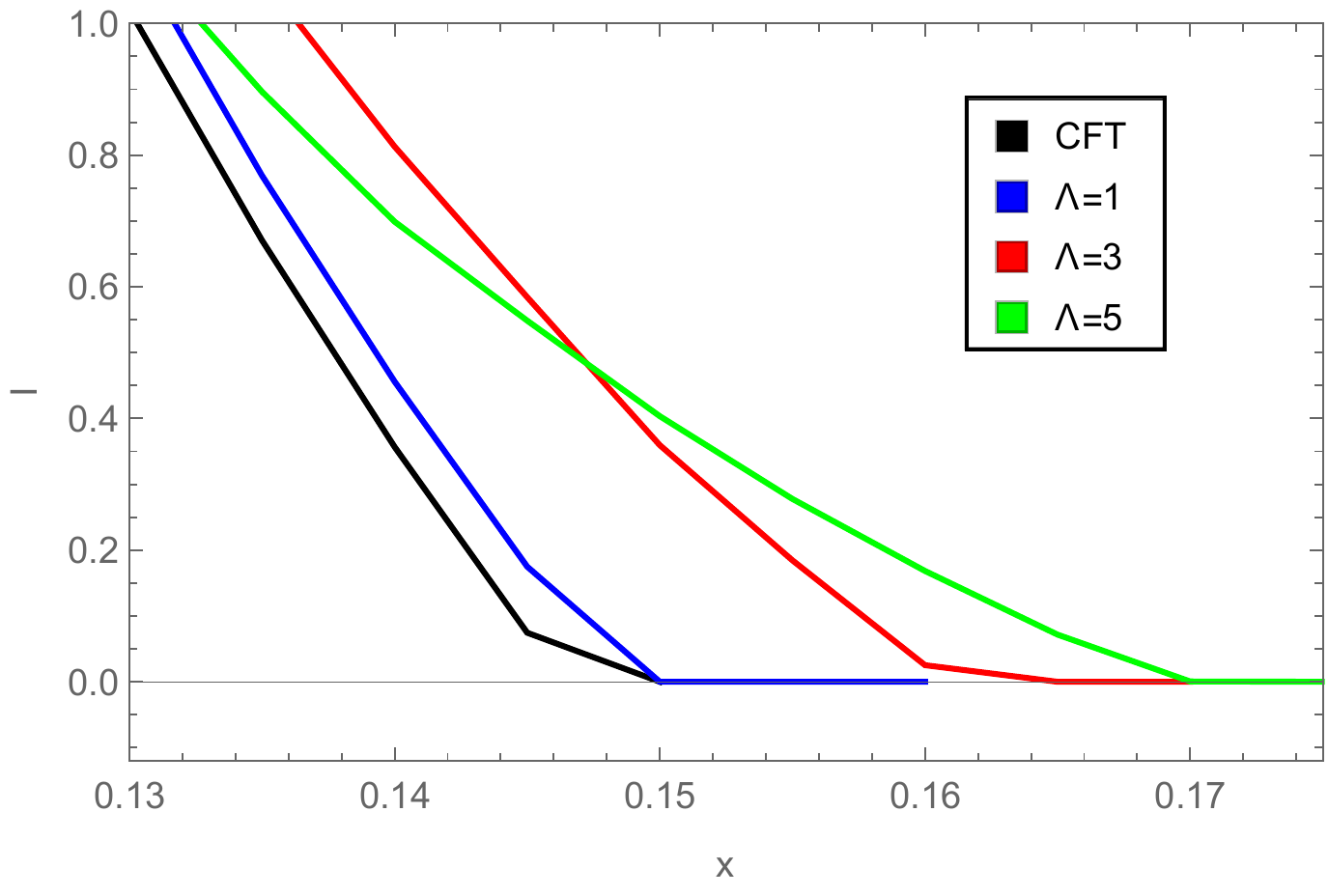}
\hspace{10mm}
\caption{The holographic mutual information of two subregions with the same length $l=0.2$ as a function of $ x $. Different curves correspond to different values of $\Lambda$ for fixed $\phi_M=10$.  The black curve corresponds to $CFT$.}\label{fig2}
\end{center}
\end{figure}
\begin{figure}[ht]
\begin{center}
\includegraphics[width=70mm]{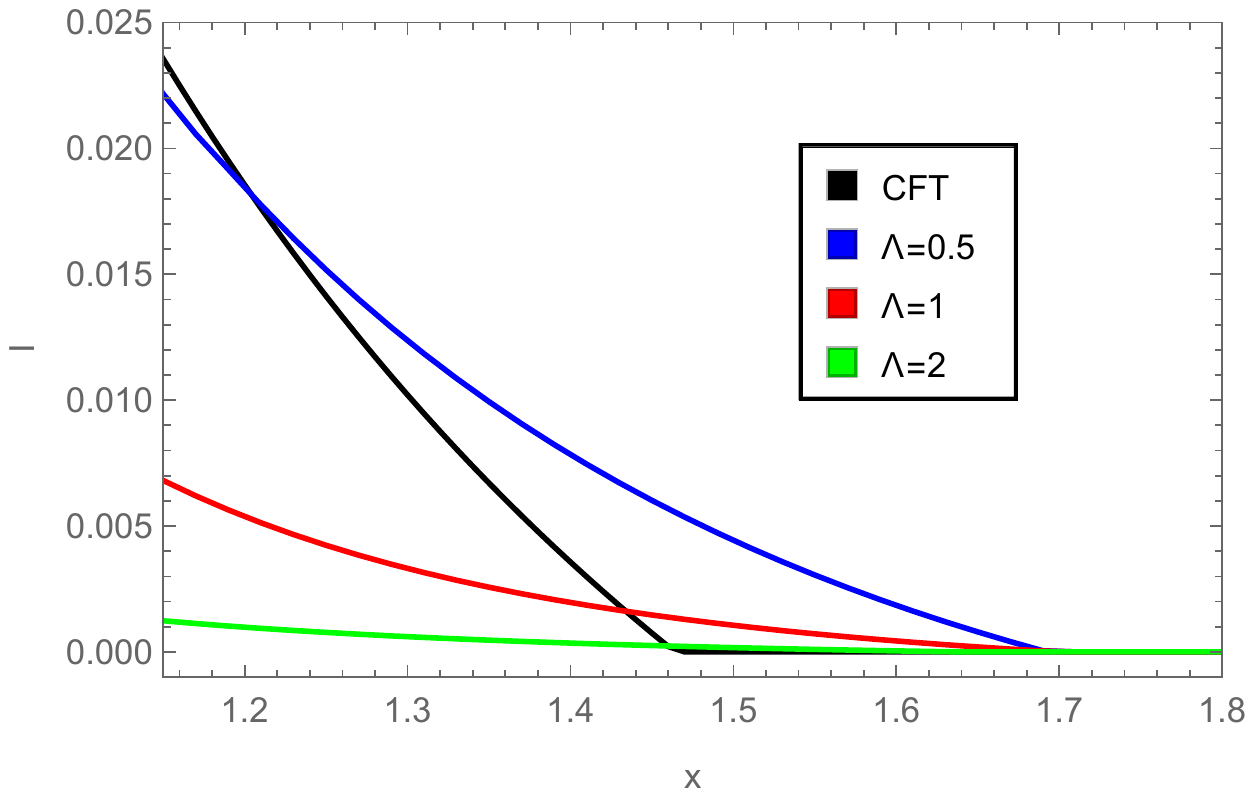}
\hspace{10mm}
\caption{The holographic mutual information of two subregions with the same length $l=2$ as a function of $ x $. Different curves correspond to different values of $\Lambda$ for fixed $\phi_M=10$.  The black curve corresponds to $CFT $.}\label{fig1}
\end{center}
\end{figure}
\section{Numerical Results}
 In the following we represent the numerical results corresponding to study the effect of  parameters such as  energy scale $\Lambda$ and the length of two subsystems $l$ on the mutual and tripartite information.

\begin{table}[ht]
\caption{$x_{DT}$ for different values of $\Lambda$ and fixed $\phi_M=10$.}
\vspace{1 mm}
\centering
\begin{tabular}{c c c c c c c c}
\hline\hline
$l$ &$\Lambda=9 $ &$ \Lambda=5 $ & $ \Lambda=3 $ &$ \Lambda=1 $ & $ \Lambda=0.5 $
 & CFT\\
 \hline\hline
\,\,\,\,\,\,\,\,\,\,\,\,\,\,\,0.2 \quad $x_{DT}$ &0.171 & 0.171 & 0.161 & 0.149 & 0.148&0.147\\
\,\,\,\,\,\,\,\,\,\,\,\,\,\,\,2 \quad $x_{DT}$ &1.566 & 1.59 &1.615& 1.70& 1.70&1.47\\
\hline
\end{tabular}\\[1ex]
\label{l02}
\end{table}
The two spatial regions $A$ and $B$  are equal intervals whose lengths are  $l=2$ and $l=0.2$, respectively. In Figs. \ref{fig2} and \ref{fig1} we have plotted the holographic mutual information as a function of $x$ for fixed $\phi_M=10$. The  black curve represents the corresponding holographic mutual information in the CFT, while the colored ones are characterized by distinct values of $ \Lambda$. On the other side,  different values of $x_{DT}$ corresponding to these figures have been listed in the table \ref{l02}.

 From the figures, in general, we can clearly observe the transition of the mutual information from positive values to zero in the range of $x<x_{DT}$ and  $x>x_{DT}$, respectively. Recall that the holographic mutual information is positive when the connected configuration is favored.  Furtheremore, it is observed that for small $l$ (in our case $l=0.2$), where the two subregions prob $UV$ regime of the  field theory under study, and small values of $\Lambda$ the behavior of the disentangling transition is the same as the conformal field theory that is $x_{DT}\simeq x_{DT}^{c}$. Intuitively, this is true since  two subregions do not feel the energy scale in this regime and become uncorrelated at the same distances like the $CFT$. On the other side, since $N\equiv (N_{UV}-N_{IR})>0$  one might guess that $x_{DT}$ in the non-conformal field theory ($NCFT$) should be smaller than $x_{DT}^{c}$ in the conformal field theory ($CFT$), for large values of $\Lambda$ .
In contrast, the results in the table \ref{l02} for $l=0.2$ and  especially for large values of $\Lambda$ indicate that $x_{DT}>x_{DT}^{c}$. Therefore, we do beleive that the effect of energy scale $\Lambda$ overcomes the effect of decrease of degrees of freedom and then the two subregions become disentangled at further distances. In fact, our intution, which states that the smaller is the number of degrees of freedom the smaller is the mutual information, is correct when  $\Lambda$ is small enough.
\begin{figure}[ht]
\begin{center}
\includegraphics[width=70mm]{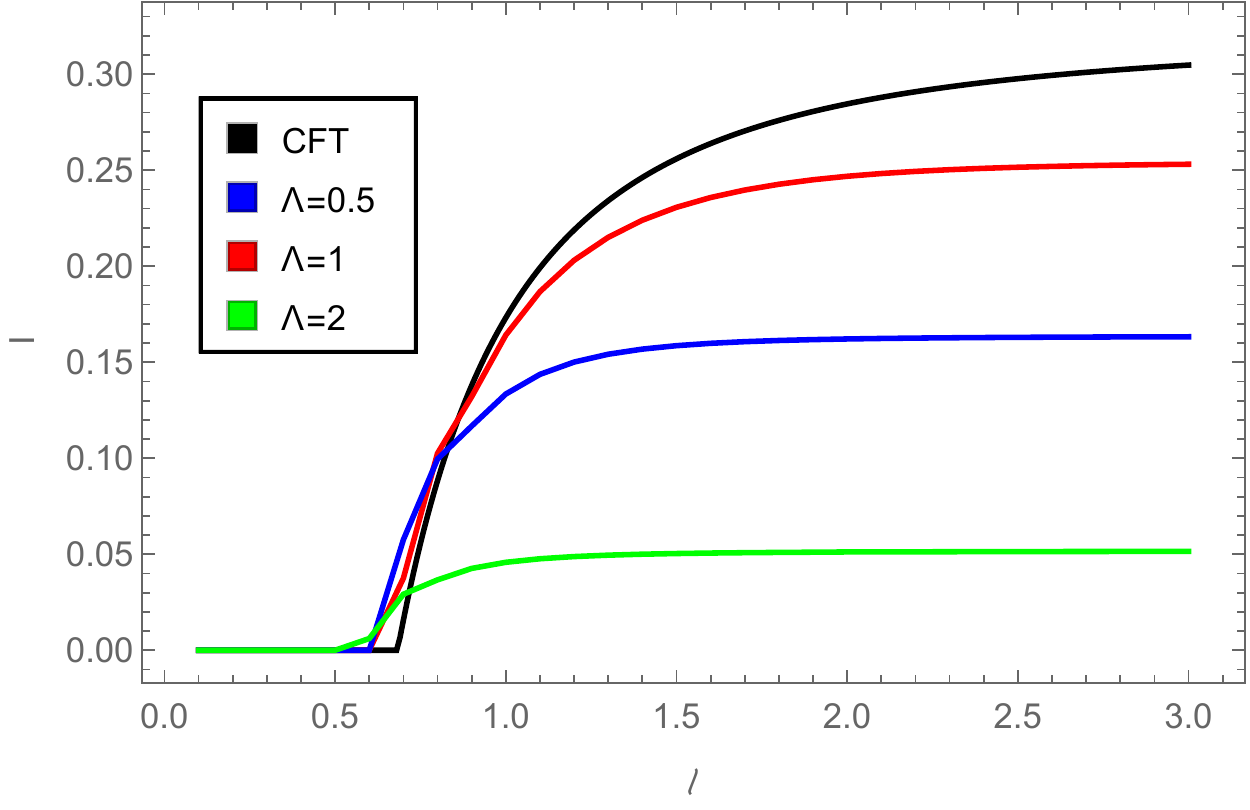}
\hspace{10mm}
\caption{The holographic mutual information of two subregions with the same separation length  $x=0.5$ as a function of $ l $. Different curves correspond to different values of $\Lambda$ for fixed $\phi_M=10$.  The black curve corresponds to $CFT $.}\label{fig111}
\end{center}
\end{figure}

  \begin{figure}[ht]
\begin{center}
\includegraphics[width=70mm]{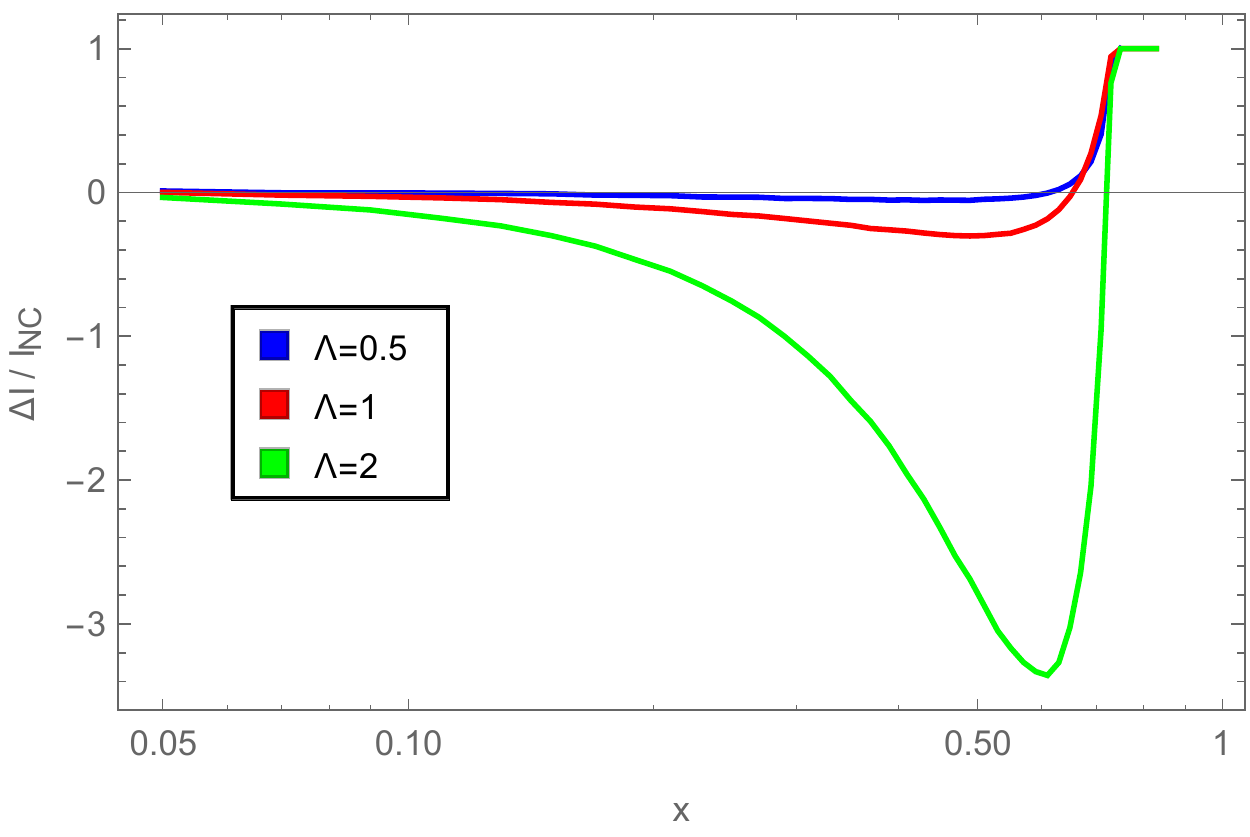}
\hspace{10mm}
\caption{$\frac{\Delta I}{I_{NCFT}}\equiv\frac{I_{NCFT} - I_{CFT}}{I_{NCFT}}$ for two subregions of the same length $l=1$ as a function of $ x $. We set $\phi_M=10$ and consider different values of $ \Lambda$.}\label{fig3}
\end{center}
\end{figure}

For large $l$ (in our case $l=2$) and large $\Lambda$ the two subregions do not meet effectively the energy scale $\Lambda$  and become uncorrelated at distances just like the $CFT$, $x_{DT}\simeq x_{DT}^{c}$. On the gravity side the turning point of the extremal surface is far away from the position of the energy scale $\Lambda$ and hence it does not change substantially the shape of this surface. Neverthelese, for small $\Lambda$ the energy scale plays the key role and $x_{DT}>x_{DT}^{c}$ which is in complete agreement with figure \ref{fig1}.\\ 

In Fig. \ref{fig111} we have shown the dependence of $I$ from the subregion size $l$. We have taken two subregions with the same length $l$ whose separation is $x = 0.5$ for fixed $\phi_M = 10$ . As one expects the transition of the mutual information from zero to positive values is occured as $l$ increases for fixed separation $x$.

We can now classify our results  as follows:
\begin{itemize}
\item The energy scale $\Lambda$ and decrease of degrees of freedom have opposite effect on the mutual information.
\item $x_{DT}$ is always bigger than $x_{DT}^{c}$ in the presence of energy scale $\Lambda$. Put it in different words, the two subregions in $NCFT$ become disentangled in larger  separation distance  than $CFT$, independent of  $l$.
\item These two regims  reveal by our numerical caculation:
 \begin{align}\label{sAB}
\begin{split}
		{l\rightarrow0 \:(E \propto l^{-1}\rightarrow\infty)}:
		\begin{cases}
			\Lambda\rightarrow\infty \quad (NCFT),  \\
			\Lambda\rightarrow 0 \,\,\,\,\quad (CFT),  \\
		\end{cases}
	\end{split}
	\end{align}
\begin{align}\label{sAB2}
\begin{split}
		{l\rightarrow\infty \:(E \propto l^{-1}\rightarrow 0)}:
		\begin{cases}
			\Lambda\rightarrow\infty \quad (CFT),  \\
			\Lambda\rightarrow 0 \,\,\,\,\quad (NCFT).  \\
		\end{cases}
	\end{split}
	\end{align}
\end{itemize}
  
In Fig. \ref{fig3}  the quantity $\frac{\Delta I}{I_{NCFT}}\equiv\frac{I_{NCFT} - I_{CFT}}{I_{NCFT}}$  has been plotted as a function of $ x $ for fixed  $\phi_M=10$. We have taken the equal subregions whose lengths are $l=1$. The main feature we notice is that by increasing $\Lambda$  the two subregions feel tangibely the appearance of the energy scale, for fixed values of $l$ and $\phi_M$. In other words, the difference between mutual information of $CFT$ and $NCFT$ become larger up to a maximum value of $x$, let's say $x_{max}$. However, this value seems mildly depend on the energy scale $\Lambda$ . Interestingly, for $x>x_{max}$, the underlying difference decreases and there is a point at which $I_{NCFT}=I_{CFT}$. This could be justified by observing that the effect of decrease of number of degrees of freedom and energy scale $\Lambda$ cancel out  eventually each other at that point. Another important point is that all curves reach one at the end of the plot since $x_{DT}> x_{DT}^{c}$ and therefore $\frac{\Delta I}{I_{NCFT}}=1$. There are also some values of $x$ where $I_{NCFT}>I_{CFT}$.\\

\begin{figure}[ht]
\begin{center}
\includegraphics[width=70mm]{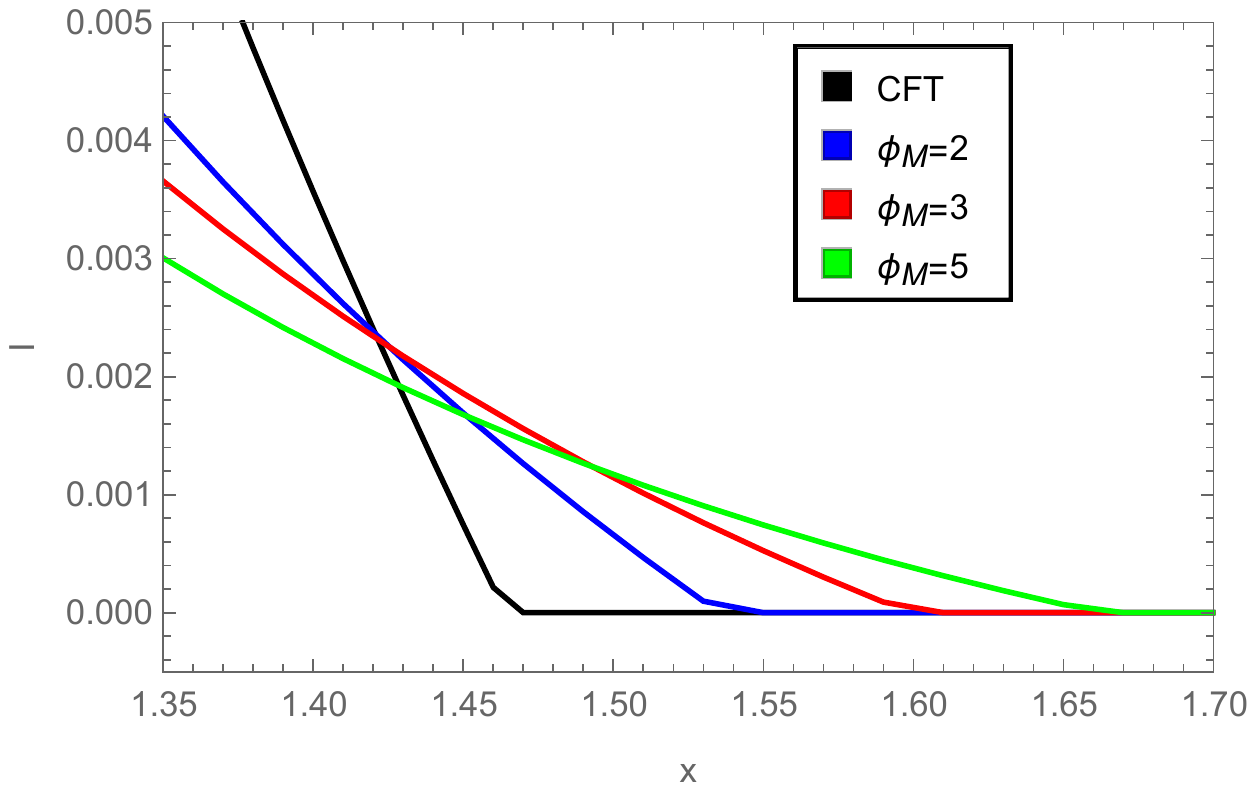}
\hspace{10mm}
\caption{The holographic mutual information $I$ of two subregions with the same length $l = 2$ as a function of the separation length $x$. We set energy scale $\Lambda=1$.  The different curves are characterized by different values of $\phi_M$.
 }\label{fig6}
\end{center}
\end{figure}

\begin{table}[ht]
\caption{$x_{DT}$ for different values of $\phi_M$ and  $\Lambda=1$.}
\vspace{1 mm}
\centering
\begin{tabular}{c c c c c c c c}
\hline\hline
$l$ &$\phi_M=8 $ &$ \phi_M=5 $ & $ \phi_M=3 $ &$ \phi_M=2 $ & $ CFT $
 & CFT\\
 \hline\hline
\,\,\,\,\,\,\,\,\,\,\,\,\,\,\,2 \quad $x_{DT}$ &1.71 & 1.67 & 1.61 & 1.55 & 1.47\\
\,\,\,\,\,\,\,\,\,\,\,\,\,\,\,8 \quad $x_{DT}$ &6.4 & 6.36 &6.25&6.05& 5.9\\
\hline
\end{tabular}\\[1ex]
\label{ll2}
\end{table}

In Fig. \ref{fig6}  the holographic mutual information as a function of $x$ has been plotted for two subregions whose lengths are $l=2$ . We have fixed $\Lambda=1$ and considered differentt values of $\phi_M$. In the table \ref{ll2} we have  listed different values of $x_{DT}$ corresponding to $l=2$ and $l=8$ as small and large length, respectively. It is followed by the table \ref{ll2} that whether $l$ is small or large the two subregions become disentangled for larger distances in the $NCFT$. In fact, by increasing  $\phi_M$ the difference in degrees of freedom between the $UV$ and $IR$ fixed points increases and hence one may expect that $x_{DT}^{c}<x_{DT}$. However, the energy scale $\Lambda$ has dominant effect and causes  $x_{DT}^{c}>x_{DT}$.  This can be seen from Fig. \ref{fig6}, Similar to the previous result.\\ 

\begin{figure}[ht]
\begin{center}
\includegraphics[width=70mm]{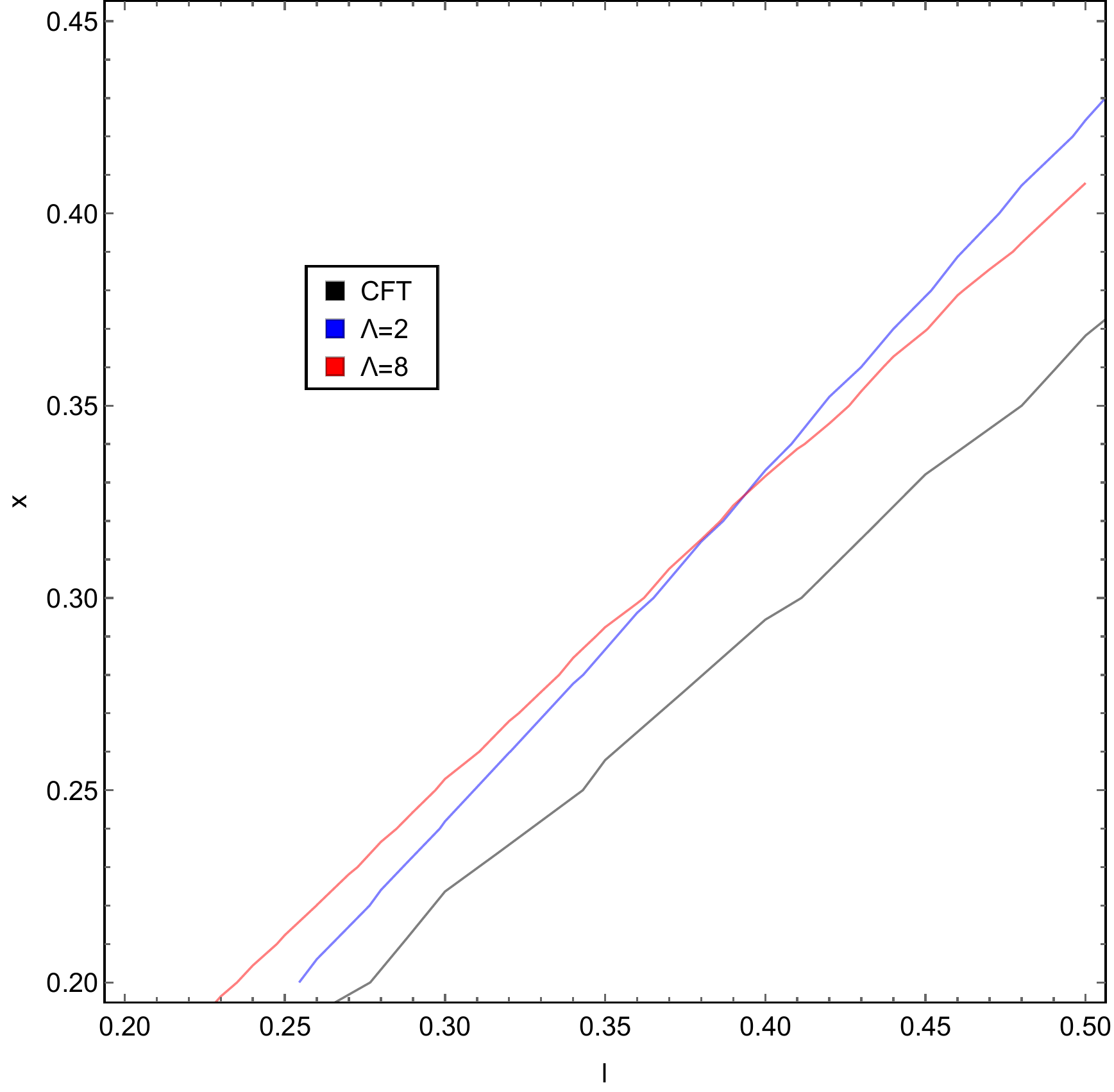}
\hspace{10mm}
\caption{The parameter space of the two subregions with the same length fixed value of $\phi_M=10$. All of the curves correspond to $I = 0$  below which the two-subsystems become entangled. Different curves correspond to different values of $\Lambda$ to study mutual information behavior.}
\label{ps1}
\end{center}
\end{figure}

\begin{figure}[ht]
\begin{center}
\includegraphics[width=70mm]{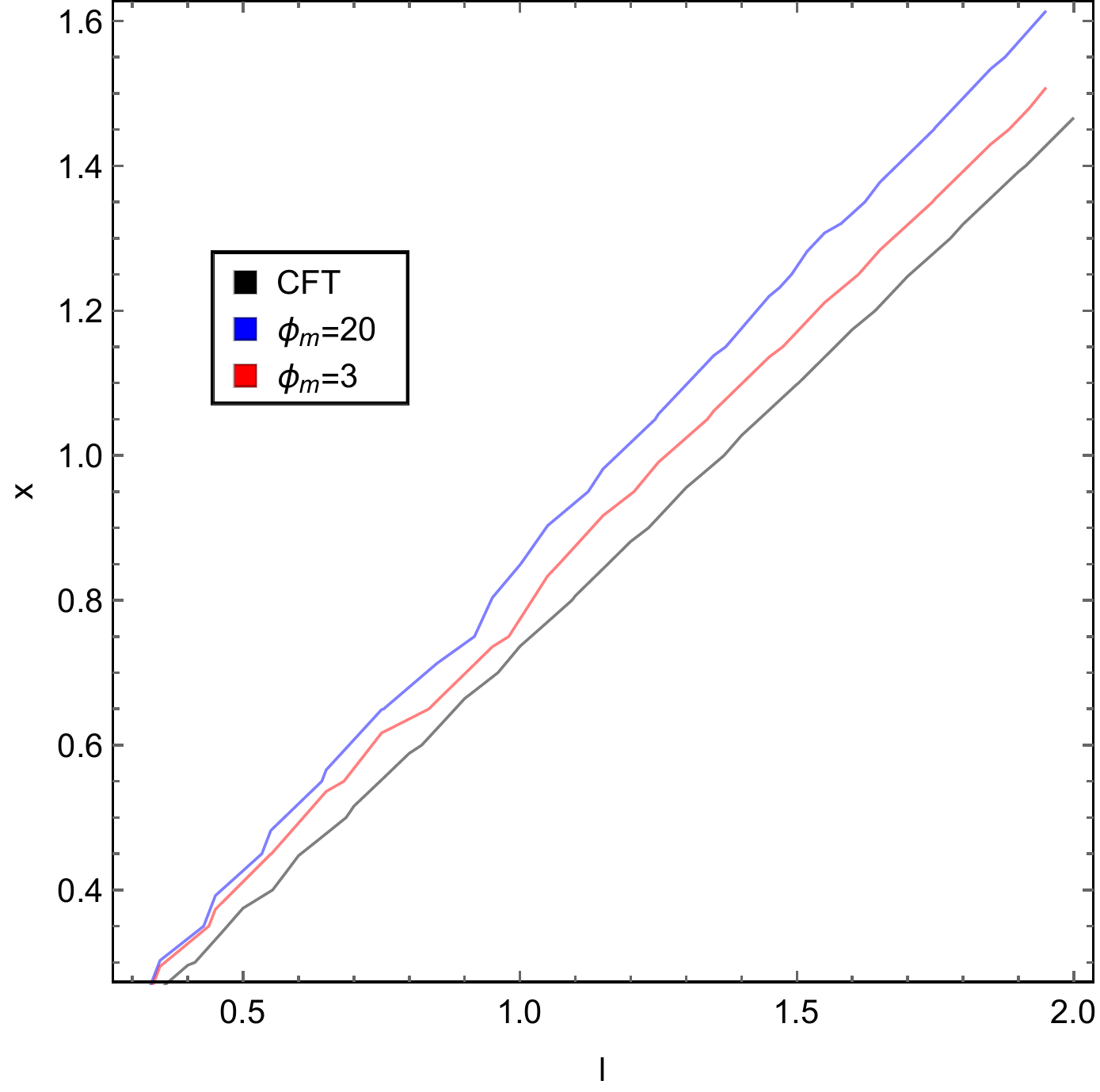}
\hspace{10mm}
\caption{The parameter space of the two subregions with the same length and fixed value of $\Lambda=2$. All of the curves correspond to $I = 0$  below which the two-subsystems become entangled. Different curves correspond to different values of $\phi_M=10$ to study mutual information behavior.}
\label{ps3}
\end{center}
\end{figure}

 In Figs. \ref{ps1} and \ref{ps3} we have pictorially shown  the phase space diagram corresponding to the two subregions with the same length $l$. In Fig. \ref{ps1} we set $\phi_M=10$ and consider different values of $\Lambda$. The diagram shows the regions where the two subregions have either non-zero or zero mutual information. All curves stand for zero mutual information and the area below them represents the regime of parameters where there is non-zero correlation between the two subregions. It can be observed from this plot that the region of the phase space where the mutual information has non vanishing value in non-conformal field theory is wider than the $CFT$'s one, coincided with the results reported in table \ref{l02}. Furtheremore, on the field theory side the non-conformal field theory behaves conformally in the $UV$ regime which is probed by very small $l$. It is evident from the figure \ref{ps1} that for small $\Lambda$ and  small values of $x$ and $l$, the the region of non-zero correlation is closer
to the conformal result compared to the large Λ, in perfect agreement with results (\ref{sAB}). On the gravity side for very small value of $l$ one may argue that the turning point of the extremal surface  gets closer to the boundary region where is asymptotically $AdS_5$  and then the deviation from this geometry's results  vanish approximatelly for the smaller $\Lambda$. This is in agreement with the table \ref{l02}. By increasing $l$ and $x$ the role of energy scale $\Lambda$ exchanges and hence the region of entanglement in the phase space becomes more limited. For smaller $\Lambda$, sufficiently deep in the $IR$, the non-zero mutual information region approaches to the $AdS_5$'s  results which is consistent with (\ref{sAB2}).
 
  In Fig. \ref{ps3} we have shown the dependence of the phase space diagram of two sub-sytems from the value $\phi_M$ for the fixed value of $\Lambda=2$. It is evident that by increasing $\phi_M$  the region of the parameter space where two subregions have non-zero correlation begins broadening. In other words, the more difference in the number of degree of freedom between $UV$ and $IR$ regime the wider region of entanglement.
   Furthermore, for small $l$ changing values of $\phi_M$ has no role and the results are approximately the same as $CFT$'s case since small values of $l$ probes the $UV$ of the field theory.

  \begin{figure}[ht]
\begin{center}
\includegraphics[width=70mm]{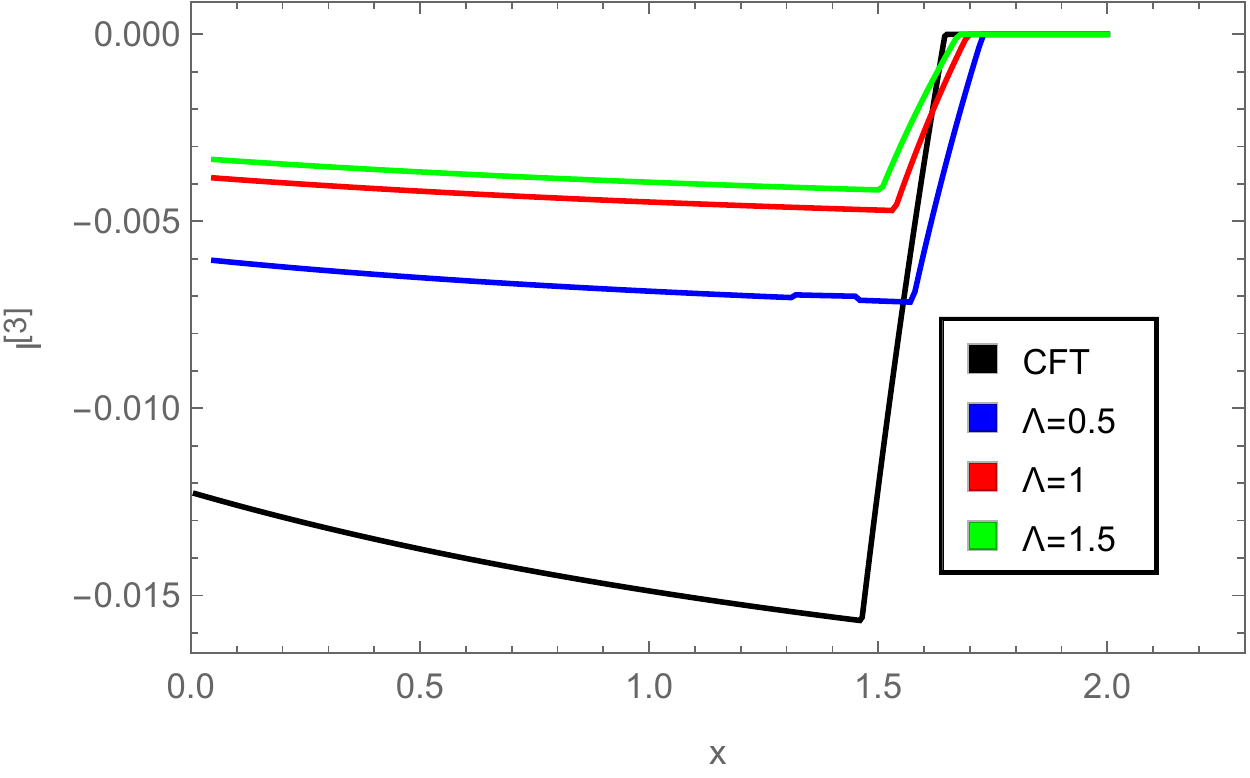}
\hspace{10mm}
\caption{The holographic tripartite information $I^{[3]}$  as a function of the the separation length $x$ for fixed length $l = 2$. We set $\phi_M=2$. The different curves are characterized by different $\Lambda$.
by different values of }\label{i3phim2}
\end{center}
\end{figure}
 \begin{figure}[ht]
\begin{center}
\includegraphics[width=70mm]{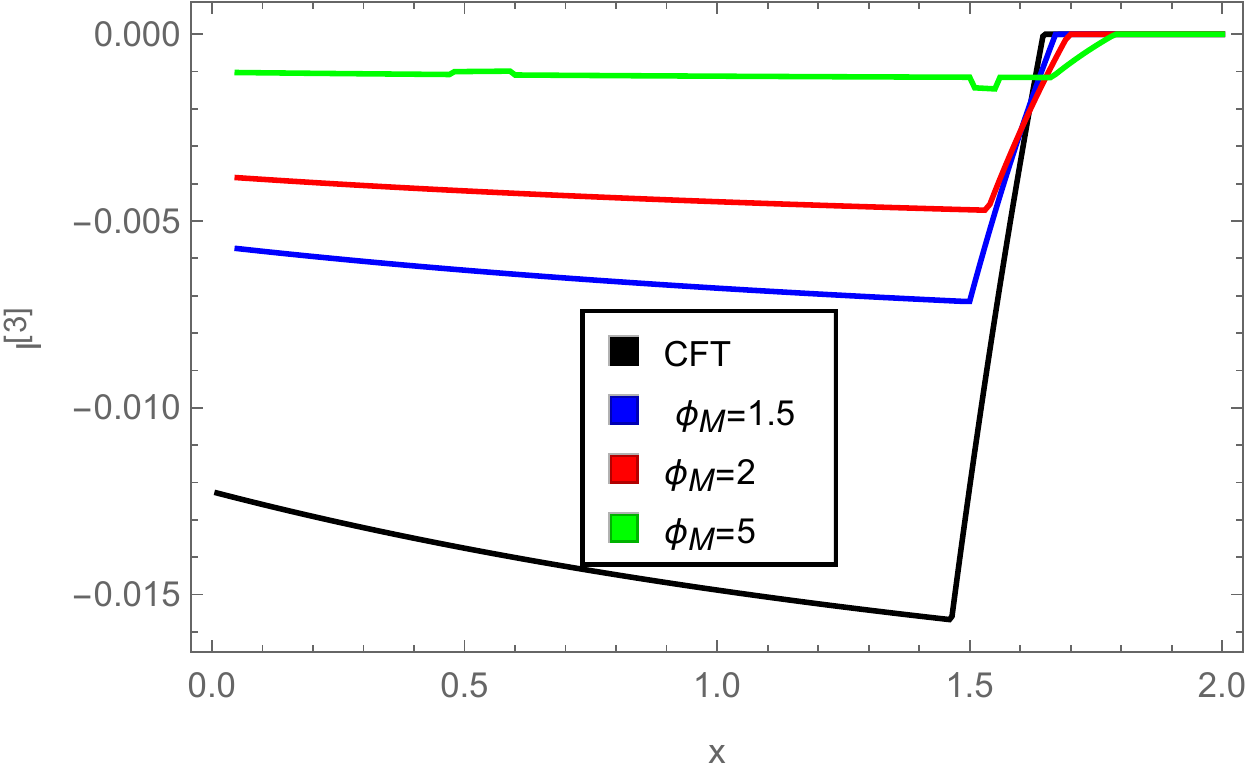}
\hspace{10mm}
\caption{The holographic tripartite information $I^{[3]}$  as a function of the the separation length $x$  for fixed length $l = 2$. We set $\Lambda=1$. The different curves are characterized by different values of $\phi_M$.}\label{i3lambda1}
\end{center}
\end{figure}

In Figs. \ref{i3phim2} and \ref{i3lambda1} we have shown the results for the tripartite information $I^{[3]}$ as a function of the distance between subregion $x$,  for different values of $\phi_M$ and $\Lambda$. In both figures the tripartite information starts at the negative values and ends at the zero (or less negative) value(s) passing through an intermediate phase where it is more negative than either. For the latter case, the initial value is more negative than the final one and the tripartite information just increases. As shown in Figs. \ref{i3phim2} and \ref{i3lambda1}, the tripartite information is generically non-positive in $NCFT$ and hence the mutual information respects also the strong subadditivity of the holographic entanglement entropy. Therefore, one can say that similar to $CFT$ the holographic mutual information is monogamous in $NCFT$. In both plots the tripartite information in
 $CFT$ is more negative than $NCFT$, independent of how one changes the values of $\Lambda$ and  $\phi_M$. From Fig. \ref{i3phim2} it is observed that  the energy scale has pushed  tripartite information towards the extensive mutual information, $i.e.$ $I^{[3]}=0$ and this process has been catalyzed by increasing $\Lambda$. On the other side, from figure \ref{i3lambda1} one can observe that increasing $\phi_M$ has the same story as the energy scale. 
\section*{Acknowledgement}
 M. A. is a post-doc fellow of Boniad Melli Nokhbegan (BMN) under Allameh Tabatabayie grant c/o M.M. Sheikh-Jabbari.


\begin{thebibliography}{99}

\bibitem{Maldacena}
J.~M.~Maldacena,
``The large N limit of superconformal field theories and supergravity,''
Adv.\ Theor.\ Math.\ Phys.\ {\bf 2} (1998) 231
[Int.\ J.\ Theor.\ Phys.\ {\bf 38} (1999) 1113][arXiv:hep-th/9711200];
S.~S.~Gubser, I.~R.~Klebanov and A.~M.~Polyakov,
``Gauge theory correlators from non-critical string theory,''
Phys.\ Lett.\ B {\bf 428} (1998) 105[arXiv:hep-th/9802109];
E.~Witten,
``Anti-de Sitter space and holography,''
Adv.\ Theor.\ Math.\ Phys.\ {\bf 2} (1998) 253[arXiv:hep-th/9802150].

\bibitem{CasalderreySolana:2011us} 
  J.~Casalderrey-Solana, H.~Liu, D.~Mateos, K.~Rajagopal and U.~A.~Wiedemann,
  ``Gauge/String Duality, Hot QCD and Heavy Ion Collisions,''
  book:Gauge/String Duality, Hot QCD and Heavy Ion Collisions. Cambridge, UK: Cambridge University Press, 2014
  [arXiv:1101.0618 [hep-th]].
  
 \bibitem{Camilo:2016kxq} 
  G.~Camilo,
  ``Expanding plasmas from Anti de Sitter black holes,''
  Eur.\ Phys.\ J.\ C {\bf 76}, no. 12, 682 (2016)
  [arXiv:1609.07116 [hep-th]].

\bibitem{Amiri-Sharifi:2016uso} 
  S.~Amiri-Sharifi, M.~Ali-Akbari, A.~Kishani-Farahani and N.~Shafie,
  ``Double Relaxation via AdS/CFT,''
  Nucl.\ Phys.\ B {\bf 909}, 778 (2016)
  [arXiv:1601.04281 [hep-th]].
  
  \bibitem{Attems:2016ugt} 
  M.~Attems, J.~Casalderrey-Solana, D.~Mateos, I.~Papadimitriou, D.~Santos-Oliván, C.~F.~Sopuerta, M.~Triana and M.~Zilhão,
  ``Thermodynamics, transport and relaxation in non-conformal theories,''
  JHEP {\bf 1610}, 155 (2016)
  [arXiv:1603.01254 [hep-th]].
  
  \bibitem{Pang:2015lka} 
  D.~W.~Pang,
  ``Corner contributions to holographic entanglement entropy in non-conformal backgrounds,''
  JHEP {\bf 1509}, 133 (2015)
  [arXiv:1506.07979 [hep-th]].
  

\bibitem{Srednicki:1993im} 
  M.~Srednicki,
  ``Entropy and area,''
  Phys.\ Rev.\ Lett.\  {\bf 71}, 666 (1993)
  [hep-th/9303048].
  
  \bibitem{Ryu:2006bv} 
  S.~Ryu and T.~Takayanagi,
  ``Holographic derivation of entanglement entropy from AdS/CFT,''
  Phys.\ Rev.\ Lett.\  {\bf 96}, 181602 (2006)
  [hep-th/0603001].
\bibitem{Ryu:2006ef} 
  S.~Ryu and T.~Takayanagi,
  ``Aspects of Holographic Entanglement Entropy,''
  JHEP {\bf 0608}, 045 (2006)
  [hep-th/0605073].
  
   
  \bibitem{Rahimi:2016bbv} 
  M.~Rahimi, M.~Ali-Akbari and M.~Lezgi,
  ``Entanglement entropy in a non-conformal background,''
  Phys.\ Lett.\ B {\bf 771}, 583 (2017)
  [arXiv:1610.01835 [hep-th]].
  
  
\bibitem{Rahimi:2018ica} 
  M.~Rahimi and M.~Ali-Akbari,
  ``Holographic Entanglement Entropy Decomposition in an Anisotropic Gauge Theory,''
  Phys.\ Rev.\ D {\bf 98}, no. 2, 026004 (2018)
  [arXiv:1803.01754 [hep-th]].
  
  \bibitem{Allais:2011ys} 
  A.~Allais and E.~Tonni,
  ``Holographic evolution of the mutual information,''
  JHEP {\bf 1201}, 102 (2012)
  [arXiv:1110.1607 [hep-th]].
  
   \bibitem{Fischler:2012uv} 
  W.~Fischler, A.~Kundu and S.~Kundu,
  ``Holographic Mutual Information at Finite Temperature,''
  Phys.\ Rev.\ D {\bf 87}, no. 12, 126012 (2013)
  [arXiv:1212.4764 [hep-th]].
  
  \bibitem{Asadi:2018ijf} 
  M.~Asadi and M.~Ali-Akbari,
  ``Holographic Mutual and Tripartite Information in a Symmetry Breaking Quench,''
  Phys.\ Lett.\ B {\bf 785}, 409 (2018)
  [arXiv:1804.05604 [hep-th]].
  
  
  \bibitem{Hayden:2011ag} 
  P.~Hayden, M.~Headrick and A.~Maloney,
  ``Holographic Mutual Information is Monogamous,''
  Phys.\ Rev.\ D {\bf 87}, no. 4, 046003 (2013)
  [arXiv:1107.2940 [hep-th]].
  
  
  
  \bibitem{Asadi:2018lzr} 
  M.~Asadi and R.~Fareghbal,
  ``Holographic Calculation of BMSFT Mutual and 3-partite Information,''
  Eur.\ Phys.\ J.\ C {\bf 78}, no. 8, 620 (2018)
  [arXiv:1802.06618 [hep-th]].
 






\bibitem{mateos16} 
  M.~Attems, J.~Casalderrey-Solana, D.~Mateos, I.~Papadimitriou, D.~Santos-Oliván, C.~F.~Sopuerta, M.~Triana and M.~Zilhão,
  ``Thermodynamics, transport and relaxation in non-conformal theories,''
[arXiv:1603.01254[hep-th]].
 


\end{thebibliography}
\end{document}